\documentclass[english]{revtex4}
\usepackage{mathptmx}

\usepackage[T1]{fontenc}
\usepackage[latin9]{inputenc}
\usepackage{amssymb}
\usepackage{graphicx}
\usepackage{esint}

\makeatletter

\usepackage{times}

\usepackage{hyperref}

\makeatother

\usepackage{babel}
\begin{document}

\title{On the mechanism for breaks in the cosmic ray spectrum}

\author{M.A. Malkov$^{1}$, P.H. Diamond$^{1,3}$, R.Z. Sagdeev$^{2}$ }

\affiliation{$^{1}$CASS and Department of Physics, University of California,
San Diego, La Jolla, CA 92093\\$^{2}$University of Maryland, College
Park, MD 20742-3280\\$^{3}$National Fusion Research Institute, Daejeon,
Korea}
\begin{abstract}
The proof of cosmic ray (CR) origin in supernova remnants (SNR) must
hinge on full consistency of the CR acceleration theory with the observations;
direct proof is impossible because of the orbit stochasticity of CR
particles. Recent observations of a number of galactic SNR strongly
support the SNR-CR connection in general and the Fermi mechanism of
CR acceleration, in particular. However, many SNR expand into weakly
ionized dense gases, and so a significant revision of the mechanism
is required to fit the data. We argue that strong ion-neutral collisions
in the remnant surrounding lead to the steepening of the energy spectrum
of accelerated particles by \emph{exactly one power}. The spectral
break is caused by a partial evanescence of Alfven waves that confine
particles to the accelerator. The gamma-ray spectrum generated in
collisions of the accelerated protons with the ambient gas is also
calculated. Using the recent Fermi spacecraft observation of the SNR
W44 as an example, we demonstrate that the parent proton spectrum
is a classical test particle power law $\propto E^{-2}$, steepening
to $E^{-3}$ at $E_{br}\approx7GeV$.
\end{abstract}
\maketitle

\section{Introduction}

The discovery of cosmic rays (CR) dates back to the historic Victor
Hess balloon ascent in 1912 \cite{Hess12}. CR origin is thus a century
old problem. Only the latest \emph{direct }observations of galactic
supernova remnants (SNRs) \cite{Enomoto02,Ahar06RXJ,Abdo10W44full,VeritasCasA10,AbdoW28_10,AbdoIC443_10,Abdo11RXJ,AgileW44_11}
narrowed the search to precisely these objects as the most probable
sources of the CRs. One serious problem on the observational side
was the lack of the SNR gamma-ray data below the energy range of the
imaging atmospheric Cerenkov telescopes, or IACT. The Fermi-LAT (large
area telescope) and Agile observatories are rapidly bridging this
gap (roughly in the 0.1-30 GeV band e.g., \cite{Abdo10W44full,AgileW44_11}),
virtually overlapping with the IACT energy band. There have been recent
breakthrough observations of such SNR as W44, IC443, W28, RX J1713
and Cas A \cite{Abdo10W44full,AbdoIC443_10,AbdoCasA10,AbdoW28_10,Abdo11RXJ}.
Overall, observations favor the diffusive shock acceleration (or DSA
\cite{Drury83,BlandEich87,MDru01}, a modern version of the mechanism
originally suggested by Fermi in 1949 \cite{Fermi49}) as a means
for the production of galactic CRs. However, there are questions,
and even some challenges, that the recent observations pose to the
theory. 

Of those, the most relevant to the proof of the SNR-CR connection
is the form of the spectrum that the theory predicts for the particular
SNR conditions. Full understanding of the spectra will allow one to
disentangle the proton (i.e., the primary CR component) emission from
a contaminating (1-2\% level) but radiatively more efficient, and
accessible to the direct observations, electron CR component. The
most recent challenge to the DSA was posed by the measurements of
the rigidity (momentum to charge) spectra of different species (most
notably proton and helium). They turned out to be different, contrary
to the DSA predictions for the ultra-relativistic rigidity range.

Note that the latter problem arose from the indirect observations
of the background CRs \cite{Adriani11,CREAM10,ATIC2_09,AMS_prot_00},
as opposed to the above mentioned direct observations of the putative
accelerators (SNR). Generally, it is impossible to trace CR back to
their accelerators because of the orbit scrambling. The proof of their
origin in SNRs can only be achieved by proving the acceleration theory
consistent with all accessible observations. It should be noted that
'direct' observations also provide only the secondary photon emission
generated by accelerated particles, either electrons (through synchrotron,
Bremsstrahlung and inverse Compton radiation), or protons (through
their collisions with the ambient gas material). Therefore, such observations
cannot be interpreted as an evidence of proton acceleration in SNR
without a detailed understanding of the emission mechanism. Note that
electron acceleration in SNRs to at least $\sim100$ TeV is held proven
``beyond a reasonable doubt'' after the observations of the SNR
1006 by ASCA and other X-ray instruments \cite{Koyama95,Allen01}.

This paper deals with the modification of the DSA proton spectra in
a partially ionized SNR environment and its signatures in gamma-emission
from such remnants. The recent discovery of the proton/helium anomaly
in the background CR spectra is discussed elsewhere (\cite{MDSPamela12},
see also \cite{VladimirMoskPamela11,Ohira11,BlasiChemComp11} for
other suggestions to explain this anomaly). Here we pursue an alternative,
complementary approach to more common multi-band treatments, e.g.,
\cite{TanakaRXJ08}, where the fits are primarily focused on the overall
agreement across the entire spectrum (from radio to gamma). By contrast,
we concentrate on the gamma-ray band and fit an important signature
of the spectrum which is the spectral break. We believe it conveys
an important information about the physics of acceleration missed
in the 'standard' DSA theory. The quality of our fit, with virtually
no adjustable parameters, should testify for the underlying physical
scenario behind the emission. The broad-band fits do not typically
meet high-quality criteria, as they seek to fit several portions of
the data simultaneously by adjusting, in some cases, a few free parameters.
Nevertheless, they provide an excellent consistency check for each
particular model.

The recent Fermi-LAT observations of the SNRs W44 and IC443 \cite{Abdo10W44full,AbdoIC443_10}
indicate that the spectrum of the gamma ray producing protons is substantially
steeper in its high energy part than the DSA predicts. A similar discrepancy
has been found in the high energy gamma ray spectra measured by e.g.,
the CANGAROO \cite{Enomoto02}, HESS \cite{Ahar06RXJ} and MAGIC \cite{MAGICW51C11}
atmospheric Cerenkov telescopes. The lack of understanding of the
primary particle spectra triggered debates about the nature of the
observed gamma-ray emission (hadronic vs leptonic), e.g., \cite{Pohl02}.
We argued \cite{MDS05} that when a SNR interacts with a dense molecular
cloud complex, the conditions for particle confinement to the shock
are different from those adopted in conventional DSA modeling. Since
the propagation of resonant Alfven waves is inhibited by ion-neutral
collisions, particles are not confined and so escape the emission
volume. These phenomena should result in a \emph{spectral break} in
the parent proton and thus, in the gamma-ray spectrum. The spectral
index at the break should change by exactly one power $\Delta q=1$
due to an effective reduction of particle momentum space dimension
by one, since particles are confined in coordinate space only when
they are within a slab in momentum space oriented perpendicular to
the local mean magnetic field. Note that the earlier HESS observations
of the SNR RXJ 1713 were also consistent with such a break \cite{Ahar06RXJ}.
The most convincing evidence for the breaks of index one, however,
provide the recent Fermi-LAT and Agile observations of W44 \cite{Abdo10W44full,Uchiyama10,AgileW44_11}
(re-analyzed in \cite{MDS_11NatCo}), the MAGIC observations of the
SNR W51C \cite{MAGICW51C11,MagicW51_12} as well as the FERMI observations
\cite{Neronov12} of giant molecular clouds (GMC), where the Alfven
wave evanescence should also result in a $\Delta q\simeq1$ steepening
of the $E^{-q}$ CR primary spectrum. These observations are encouraging
in that they unambiguously confirm the breaks. However, they rule
out traditional DSA models based on a single power law with an exponential
cutoff.

\section{Mechanism for the spectral break}

The physics of the spectral break considered here is very simple.
When a SNR shock approaches a molecular cloud (MC) or a pre-supernova
swept-up shell, confinement of accelerated particles deteriorates.
Due to the particle interaction with magnetic turbulence, confinement
generally requires scales similar to the particle gyroradius \cite{Drury83,BlandEich87}.
However, strong ion-neutral collisions substantially enhance the role
of particle pitch angle in wave-particle interaction. While the waves
are in a strongly ionized (closer to the shock) medium they propagate
freely in a broad frequency range at the Alfven speed $V_{A}=B/\sqrt{4\pi\rho_{i}}$
with the frequencies $\omega=kV_{A}$. Here $k$ is the wave number
(assumed parallel to the local field $\mathbf{B}$) and $\rho_{i}$
is the ion mass density. As long as the Alfven wave frequency is higher
than the ion-neutral collision frequency $\nu_{in}$, the waves are
weakly damped. When, on the other hand, the ion-neutral collision
frequency is higher (deeper into the cloud), neutrals are entrained
by the oscillating plasma and the Alfven waves are also able to propagate,
albeit with a factor $\sqrt{\rho_{i}/\rho_{0}}<1$ lower speed, where
$\rho_{0}$ is the neutral density. The propagation speed reduction
occurs because every ion is now ``loaded'' with $\rho_{0}/\rho_{i}$
neutrals. In between these two regimes Alfven waves are heavily damped
and even disappear altogether for sufficiently small $\rho_{i}/\rho_{0}\ll0.1$.
The evanescence wave number range is then bounded by $k_{1}=\nu_{in}/2V_{A}$
and $k_{2}=2\sqrt{\rho_{i}/\rho_{0}}\nu_{in}/V_{A}$. These phenomena
have been studied in detail in \cite{KulsrNeutr69,ZweibelShull82},
and specifically in the context of the DSA in \cite{VoelkNeutrDamp81,DruryNeutral96,RevilleNeutr08}.
Now we turn to their impact on the particle confinement and emissivity.

\begin{figure}
\includegraphics[clip]{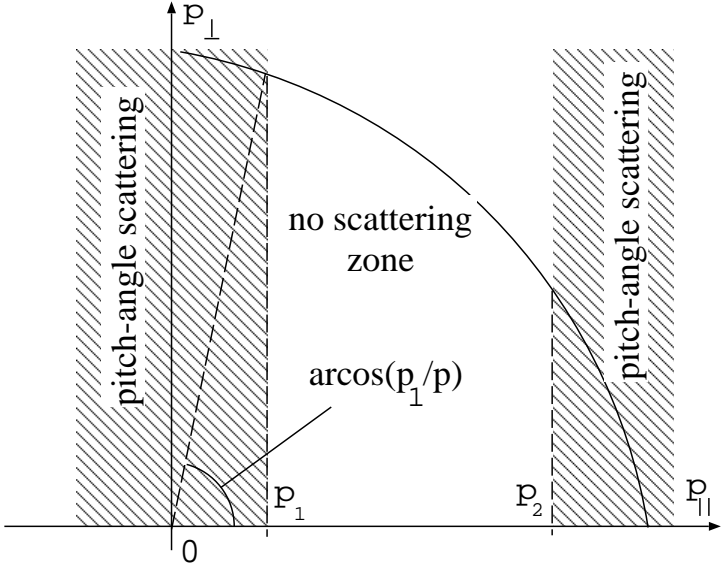}

\caption{\textbf{Momentum space of accelerated protons. }Particle scattering
zones on the $\left(p_{\parallel},p_{\perp}\right)$- plane of momentum
space. Protons in the stripes $p_{1}<\left|p_{\parallel}\right|<p_{2}$
are not scattered by waves (see text). Therefore, particles from the
domains $\left|p_{\parallel}\right|>p_{2}$ maintain their propagation
direction and promptly escape from the dense gas region. \label{fig:Scattering-zone-of}}
\end{figure}

In the frame work of a quasilinear wave-particle interaction the wave
number $k$ is approximately related to the parallel (to the magnetic
field) component of the particle momentum $p_{\parallel}$ by the
cyclotron resonance condition $kp_{\parallel}/m=\pm\omega_{c}$, where
the (non-relativistic) gyro-frequency $\omega_{c}=eB/mc$. Note that
the appearance of $p_{\parallel}=p\mu$, where $\mu$ is the cosine
of the pitch angle (see Fig.\ref{fig:Scattering-zone-of}), instead
of the often used ``sharpened'' \cite{Skill75a} resonance condition
$kp/m=\pm\omega_{c}$ is absolutely critical for the break mechanism. 

The frequency range where the waves cannot propagate may be conveniently
translated into the parallel momentum range

\begin{equation}
p_{1}<\left|p_{\parallel}\right|<p_{2},\label{eq:ineq}
\end{equation}
with 

\begin{equation}
p_{1}=2V_{A}m\omega_{c}/\nu_{in},\;\; p_{2}=\frac{p_{1}}{4}\sqrt{\rho_{0}/\rho_{i}}>p_{1}.\label{eq:p12}
\end{equation}
That a spectral break must form at the photon energy corresponding
to the particle momentum $p=p_{1}=p_{br}$, can be readily understood
from Fig.\ref{fig:Scattering-zone-of}. The 'dead zones' $p_{1}<\left|p_{\parallel}\right|<p_{2}$
imply that particles with $\left|p_{\parallel}\right|>p_{1}$ do not
turn around (while moving along the magnetic field) and escape from
the region of CR-dense gas collisions at a $p_{\parallel}/p$ fraction
of the speed of light. More specifically, particles with $p_{1}<\left|p_{\parallel}\right|<p_{2}$
escape because they are not scattered, whereas particles with $\left|p_{\parallel}\right|>p_{2}$,
while being scattered, maintain the sign of $p_{\parallel}$, as they
cannot jump over the gap $p_{1}<\left|p_{\parallel}\right|<p_{2}$
and so escape as well. An exception to this are particles with sufficiently
large $p_{\perp}$ that can be mirrored across the gap or overcome
it via the resonance broadening. We return to this possibility later. 

\begin{figure}
\includegraphics[clip,angle=-90,scale=0.5]{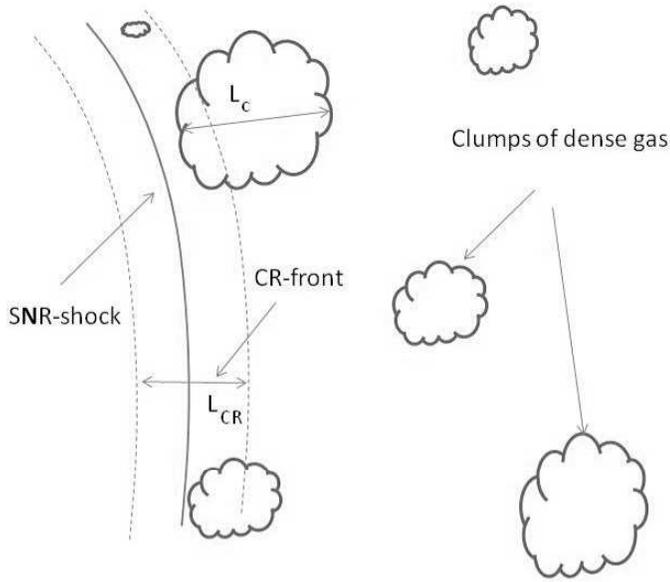}

\caption{\textbf{SNR shock propagating into dense gas environment}. The filling
factor of the gas clumps is small, while some of them may be larger
than the thickness of the CR layer near the shock front.\label{fig:SNR-shock-propagating} }
\end{figure}

The break can also be explained in terms of the confinement times
of different groups of particles introduced above. Particles with
$\left|p_{\parallel}\right|>p_{1}$ spend only short time $\tau_{{\rm esc}}\sim L_{{\rm c}}/c$
(where $L_{{\rm c}}$ is the size of the clump) inside the gas clumps.
They propagate ballistically and their scattering time is assumed
to be infinite, as there are no waves they can interact with resonantly
($p_{1}<\left|p_{\parallel}\right|<p_{2}$) or they cannot change
their propagation direction ($\left|p_{\parallel}\right|>p_{2}$).
Particles with $\left|p_{\parallel}\right|<p_{1}$ are, on the contrary,
scattered intensively in pitch angle, they frequently change their
direction, and so sit in the clump for $\tau_{{\rm conf}}\sim L_{{\rm c}}^{2}/\kappa\sim L_{{\rm c}}^{2}/c^{2}\tau_{{\rm sc}}$.
Here $\tau_{{\rm sc}}$ is their pitch-angle scattering time and $\kappa$
is the associated diffusion coefficient. Not only $\tau_{{\rm conf}}\gg\tau_{{\rm esc}}$
is required, i.e., $\tau_{{\rm sc}}\ll L_{{\rm c}}/c$, but also $\tau_{{\rm conf}}>L_{{\rm c}}/U_{{\rm sh}}$,
which means that the shock precursor is shorter than the clump $\kappa/U_{{\rm sh}}\lesssim L_{{\rm CR}}<L_{{\rm c}}$
(here $U_{{\rm sh}}$ is the shock velocity, and $L_{{\rm CR}}$ is
the thickness of the CR front near the shock). The last condition
ensures that particles with $p_{\parallel}>p_{1}$ that escape through
the clump after having entered it from the shock side, will not interact
with the shock after they exit through the opposite side of the clump,
thus escaping upstream, Fig.\ref{fig:SNR-shock-propagating}. The
reason for that is a low level of Alfven wave turbulence ahead of
the CR precursor. We also assume that the ambient magnetic field does
not deviate strongly from the shock normal, in order to allow these
particles to escape through the far side of the clump.

While particles with $p>p_{1}$ escape from the regions of enhanced
gamma radiation (high gas density), an initially isotropic distribution
of accelerated particles is maintained only in a slab in momentum
space $\left|p_{\parallel}\right|<p_{1}$ and becomes thus highly
anisotropic (a 'pancake' distribution). What matters for the integral
emission, however, is a locally isotropic component $\overline{f}$
of this new proton distribution. It can be introduced by re-averaging
the 'pancake' ($\left|p_{\parallel}\right|<p_{1}$) distribution in
pitch angle, $\overline{f}\left(p\right)\equiv\intop_{0}^{1}f\left(p,\mu\right)d\mu$,
and is readily obtained assuming that particles remaining in the dense
gas (those with $\left|p_{\parallel}\right|<p_{1}$) maintain the
flat pitch-angle distribution, i.e. 

\begin{eqnarray}
\overline{f}\left(p\right) & = & \intop_{0}^{\mu_{1}}f_{0}\left(p\right)d\mu=\left\{ \begin{array}{cc}
\left(p_{1}/p\right)f_{0}\left(p\right), & p\ge p_{1}\\
f_{0}\left(p\right), & p<p_{1}
\end{array}\right.\label{eq:fbar}
\end{eqnarray}
where $f_{0}\left(p\right)$ is the initial (isotropic) distribution
function and $\mu_{1}=\min\left\{ p_{1}/p,1\right\} $. Thus, the
slope of the particle momentum distribution becomes steeper by exactly
one power above $p=p_{1}\equiv p_{{\rm br}}$. In particular, any
power-law distribution $\propto p^{-q}$, upon entering an MC, turns
into $p^{-q-1}$ at $p\ge p_{{\rm br}}$, and preserves its form at
$p<p_{{\rm br}}$.

Note that the broken power-law spectrum can only be maintained if
the filling factor $f_{{\rm gas}}$ of the dense gas with the significant
wave evanescence interval $\left(p_{1},p_{2}\right)$ is relatively
small, $f_{{\rm gas}}\ll1$, so that the overall particle confinement
and thus the acceleration are not strongly affected. If, on the contrary,
$f_{{\rm gas}}\sim1$, the resonant particles would leak into the
$(p_{1},p_{2})$ gap and escape from the accelerator in large amounts,
thus suppressing the acceleration. We discuss further limitations
of the mechanism in Sec.\ref{sec:Summary-and-outlook} but we note
here that particles with sufficiently high momenta $p>p_{2}B_{0}/\delta B$,
where $\delta B/B_{0}$ is the effective mirror ratio of magnetic
perturbations, can ``jump'' over the gap. The primary $p^{-q}$
slope should then be restored for such particles. Recent MAGIC observations
of the SNR W51C \cite{MAGICW51C11,MagicW51_12} indeed point at such
spectrum recovery at higher energies. It should also be noted, that
the $\Delta q=1$ break index is a limiting case of identical gas
clumps. The integrated emission from an ensemble of clumps with different
$p_{1}$ and $p_{2}$ should result in a more complex spectrum.

\begin{figure}
\includegraphics[bb=0bp 0bp 792bp 612bp,clip,scale=0.5,angle=-90]{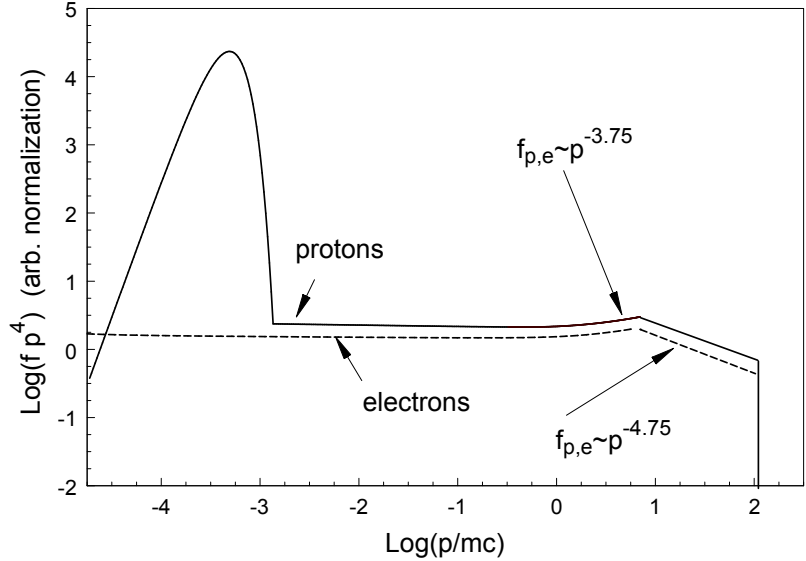}

\caption{\textbf{Spectra of accelerated protons and electrons. }The both particle
distributions are calculated for a weakly modified shock and are shown
in momentum normalization ($f\left(p\right)$ is steeper by two powers
than the spectra in energy normalization, used in the text). Both
spectra are multiplied by $p^{4}$, so that the test particle distribution
is flat. Shock parameters: acoustic Mach number $M=30$, shock velocity
$V_{s}/c=10^{-3}$, the break momentum $p_{br}\simeq7mc$. Shock pre-compression
(flow compression across the CR precursor) R=1.8, injection parameter
$\nu\simeq0.1$ {[}defined as $\nu=\left(4\pi/3\right)\left(mc^{2}/\rho V_{s}^{2}\right)\left(p_{inj}/mc\right)^{4}f\left(p_{inj}\right)$,
with $\rho$ and $V_{s}$ being the ambient gas density and the shock
speed, respectively{]}; injection momentum $p_{inj}/mc\simeq1.4\cdot10^{-3}$.\label{fig:Spectra-of-protons}}
\end{figure}

\section{Break momentum}

While the one power spectral break in the pitch-angle averaged particle
distribution seems to be a robust environmental signature of a weakly
ionized medium into which the accelerated particles propagate, the
break momentum remains uncertain. According to eq.(\ref{eq:p12}),
$p_{br}$ ($\equiv p_{1}$) depends on the magnetic field strength
and ion density as well as on the frequency of ion-neutral collisions,
$\nu_{in}=n_{0}\left\langle \sigma V\right\rangle $. Here $\left\langle \sigma V\right\rangle $
is the product of the collision cross-section and collision velocity
averaged over the thermal distribution. Using an approximation of
\cite{DraineMcKee93,DruryNeutral96} for $\left\langle \sigma V\right\rangle $,
$p_{br}$ can be estimated as 
\begin{equation}
p_{br}/mc\simeq10B_{\mu}^{2}T_{4}^{-0.4}n_{0}^{-1}n_{i}^{-1/2}.\label{eq:p1}
\end{equation}
Here the gas temperature $T_{4}$ is measured in the units of $10^{4}K$,
magnetic field $B_{\mu}$ -in microgauss, $n_{0}$ and $n_{i}$ (number
densities corresponding to the neutral/ion mass densities $\rho_{0}$
and $\rho_{i}$) \--in $cm^{-3}$. Note that the numerical coefficient
in the last expression may vary depending on the average ion and neutral
masses and can be higher by a factor of a few \cite{KulsrNeutr69,NakanoMC84}
than the estimate in eq.(\ref{eq:p1}) suggests. The remaining quantities
in the last formula are also known too poorly to make an accurate
independent prediction of the position of the break in the gamma ray
emission region. Those are the regions near the blast wave where complicated
physical processes unfold. They include particle acceleration, strong
MHD turbulence (driven by particles and their interaction with ambient
gas inhomogeneities), gas ionization by shock generated UV photons,
turbulent plasma heating and even evaporation of magnetic cloudlets
\cite{ShullMcKeeMC79,DraineMcKee93,BykovMC00}. Also important may
be the ionization by the low energy CRs accelerated at the blast wave.
However, as their diffusion length is shorter than that of the particles
with $p\gtrsim p_{br}$, we may assume that they do not reach the
MC. Pre-ionization by the UV photons can also be ignored for the column
density $N>10^{19}cm^{-2}$ ahead of the shock beyond which they are
absorbed \cite{Uchiyama10}. The authors or the Ref. \cite{Uchiyama10},
using the earlier data from \cite{Reach05} have also analyzed the
parameters involved in eq.(\ref{eq:p1}) and found the above estimate
of $p_{br}$ to be in a good agreement with the spectral break position
measured by the \emph{Fermi} LAT. Nevertheless, we may run the argument
in reverse and use the \emph{Fermi} observations \cite{Abdo10W44full}
of the gamma-ray spectrum of SNR W44 to determine the break momentum
in the parent particle spectrum and constrain the parameters in eq.(\ref{eq:p1}).
Since we also know the amount of the slope variation $\Delta q$,
we can calculate the full spectrum up to the cut-off energy.

It should also be noted that in reality the break at $p=p_{{\rm br}}$
is not infinitely sharp for the following reasons. The break momentum
may change in space due to variations of the gas parameters (eq.{[}\ref{eq:p1}{]}),
the resonance broadening \cite{Dupree66,Acht81a} near $p=p_{1}=p_{{\rm br}}$
(so that particles with $p\gtrsim p_{1}$ are still scattered, albeit
weakly) and other factors, such as the contribution of small gas clumps
with $L_{c}\ll L_{{\rm CR}}$, Fig.\ref{fig:SNR-shock-propagating}.
The small clumps are submerged in the CR front and the CRs that escape
from them are readily replenished. Note that this effect may decrease
the break index $\Delta q$. However, the conversion of the parent
proton spectrum into the observable gamma emission introduces a significant
smoothing of the break, so that even a sharply broken proton spectrum
produces a smooth gamma spectrum. It provides an excellent fit to
the Fermi gamma data without an ad hoc proton break smoothing adopted
by the Fermi-team \cite{Abdo10W44full} to fit the data. This will
be seen from our calculation of the gamma emission based on the sharp
proton spectral break in Sec.\ref{sec:Photon-spectra} below.

\section{Particle spectra}

To calculate the particle spectra, we need to determine the degree
of nonlinear modification of the shock structure. In principle, it
can be calculated consistently, given the shock parameters and the
particle maximum momentum, $p_{max}$. In the case of a broken spectrum,
$p_{br}$ likely plays the role of $p_{max}$, as a momentum where
the dominant contribution to the pressure of accelerated particles
comes from, thus setting the scale of the modified shock precursor.
Note that in the conventional nonlinear (NL) acceleration theory,
the cut-off momentum $p_{max}$ plays this role, because the nonlinear
spectra are sufficiently flat so as to make the pressure diverge with
momentum, unlike broken spectra.

The break in the photon spectrum is observed at about $2$ GeV, which
places the break in the proton distribution at about $p_{br}\simeq7GeV/c$
\cite{Abdo10W44full}. For the strength of the break $\Delta q=1$,
the spectrum above it is clearly pressure converging, and we perform
the calculation of the shock structure and the spectrum using this
break momentum as the point of the maximum in the CR partial pressure.
Note that outside of gas clumps the CR pressure may still come from
higher momenta and the complete nonlinear calculation of the spectrum
would require the filling factor of the gas clumps. However, once
the break momentum is set, we can use an analytic approach \cite{m97a,MDru01}
for a stationary nonlinear acceleration problem using $p_{br}$ as
an input parameter. 

Apart from $p_{br}$, the nonlinear solution depends on a number of
other parameters, such as the injection rate of thermal particles
into acceleration, Mach number, the precursor heating rate and the
shock velocity $V_{s}$. Of these parameters the latter is known reasonably
well, $V_{s}\approx300km/s$, the injection rate can be either calculated
analytically for the parallel shock geometry \cite{mv95,m98}, or
inferred from the simulations \cite{SpitkovskyHybr12}, while the
other parameters are still difficult to ascertain. Fortunately, in
sufficiently strong shocks the solution either stays close to the
test particle (TP) solution (leaving the shock structure only weakly
modified) or else it transitions to a strongly modified NL-solution
regime. The TP regime typically establishes in cases of moderate Mach
numbers, low injection rates and low $p_{max}$ (now probably closer
to $p_{br}$), while the NL regime is unavoidable in the opposite
part of the parameter space.

In the TP regime the spectrum is close to a power-law with the spectral
index 2 throughout the supra-thermal energy range. In the NL regime,
however, the spectrum develops a concave form, starting from a softer
spectrum at the injection energy, with the index $q\simeq(r_{s}+2)/(r_{s}-1)>2$,
where $r_{s}<4$ is the sub-shock compression ratio. Then it hardens,
primarily in the region $p\sim mc$, where both the partial pressure
and diffusivity of protons change their momentum dependence. The slope
reaches its minimum at the cut-off (break) energy, which, depending
on the degree of nonlinearity, can be as low as 1.5 or even somewhat
lower if the cut-off is abrupt. The question now is into which of
these two categories the W44 spectrum falls? Generally, in cases of
low maximum (or, equivalently, low spectral break $p_{br}\lesssim10$)
momentum, the shock modification is weak, so the spectrum is more
likely to be in a slightly nonlinear, almost TP regime. On the other
hand, there is a putative indication from the electron radio emission
that their spectrum may be close to $q_{e}\approx1.75$, which could
be the signature of a moderately nonlinear acceleration process. It
should be remembered, however, that this is a global index across
the W44 remnant. There are resolved bright filaments where a canonical
$\alpha=-0.5$ spectrum, corresponding precisely to the TP parent
electron spectrum with $q_{e}=2$ is observed \cite{W44Radio07}.
Moreover, there are regions with the positive indices $\alpha\lesssim0.4$
which cannot be indicative of a DSA process without corrections for
subsequent spectral transformations such as an absorption by thermal
electrons \cite{W44Radio07}. These regions can very well contribute
to the overall spectral hardening discussed above, mimicking the acceleration
nonlinearity. Finally, secondary electrons give rise to the flattening
of the radio spectrum as well \cite{Uchiyama10}.

If the accelerated protons and electrons respond to the turbulence
similarly, which is almost certainly the case in the ultra-relativistic
regime, their spectra should have similar slopes there (as long as
the synchrotron losses are ignorable). In using the electron radio
spectrum as a probe for the level of acceleration nonlinearity, the
following two relations are useful. First, there is a relation between
the electron energy and the radio frequency $\nu_{MHz}=4.6\cdot B_{\mu}E_{GeV}^{2}$.
The second, already mentioned relation, $q_{e}=1-2\alpha$, links
the spectral index of radio emission $\alpha$ (assuming the radio
flux $\propto\nu^{\alpha}$) and the spectral index of the parent
electrons $q_{e}$ (assuming their energy spectrum $\propto E^{-q_{e}}$).
Once the global radio spectral index of W44, $\alpha\simeq-0.37$
\cite{W44Radio07} is generated by freshly accelerated electrons in
the frequency range $74<\nu<10700$ MHz, the electrons should maintain
their modified spectrum over the energy range spanning more than one
order of magnitude. For example, assuming $B_{\mu}\simeq70$ \cite{Abdo10W44full},
one sees that electrons must maintain an index $q_{e}\approx1.75$
between $0.46<E<5.8$ GeV. While the upper bound is acceptable given
the spectral break proton energy inferred from the super GeV emission
measured by the Fermi LAT, the lower end is rather uncomfortable,
since the nonlinear hardening of both protons and electrons with the
Bohm (or other similar for protons and electrons turbulent diffusivities)
starts (slowly) only at the proton rest energy. The calculated nonlinear
spectra are shown in Fig.\ref{fig:Spectra-of-protons} for the both
species. At and below $1$GeV, the electron spectrum stays close to
the test particle solution, $q_{e}\approx2,$ even though the proton
spectrum may steepen there, as we mentioned above. The physical reason
for this difference is that the electron mean free path falls off
slower with decreasing momentum than that of the protons in the Mev-Gev
momentum range so that electrons sample longer parts of the shock
precursor with higher flow compression and thus develop a harder spectrum.

\begin{figure}
\includegraphics[bb=0bp 0bp 792bp 612bp,scale=0.5,angle=-90]{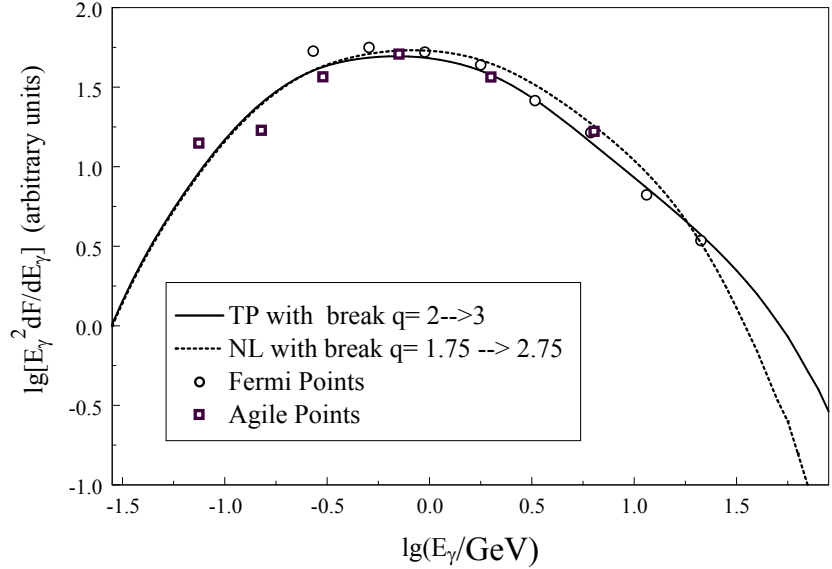}

\caption{\textbf{Gamma radiation spectra.} Photon spectra resulting from $\pi^{0}$
decay and calculated for two different parent proton spectra compared
against the Fermi (circles) and Agile (squares) data. Solid line:
a test particle acceleration regime with the spectral index $q=2$
below the break and $q=3$ above the break at $p_{br}=7GeV$/c. Dashed
line: a moderately nonlinear acceleration regime corresponding to
the spectrum shown in Fig.\ref{fig:Spectra-of-protons} ($q\simeq1.75$
and $q\simeq2.75$ below and above the break, respectively). Cut-offs
are placed at 300 GeV for TP- and 100 GeV, for NL-spectrum. Fermi
and Agile data are adopted from \cite{Abdo10W44full,AgileW44_11},
respectively. Both curves are well within the error bars of Fermi
and Agile (not shown for clarity), which, in turn, overlap \cite{AgileW44_11}.
\label{fig:Gamma-emission-from}}
\end{figure}

\section{Photon spectra\label{sec:Photon-spectra}}

The above considerations somewhat weaken the radio data as a probe
for the slope of the electron and (more importantly) for the proton
spectrum. Therefore, the exact degree of nonlinearity of the acceleration
remains unknown and we consider both the TP and weakly NL regimes
in our calculations of the photon spectra, generated in $p-p$ collisions.
Specifically, we calculate the $\pi^{0}$ production rate and the
gamma-ray emissivity. In so doing, we adopt numerical recipe described
in detail in \cite{Kamae06,KarlssonKamae08}. The physical processes
behind these calculations are (i) collisions of accelerated protons
with the protons of the ambient gas resulting in the following spectrum
of $\pi^{0}$-mesons:

\[
F_{pp}\left(E_{\pi}\right)=4\pi N_{pg}\intop\frac{d\sigma\left(E_{\pi},E_{p}\right)}{dE_{\pi}}J_{p}\left(E_{p}\right)dE_{p}
\]
where, $N_{pg}$ is the number density of protons in the gas, $d\sigma/dE_{\pi}$
is the differential cross section for the $\pi^{0}$ production in
collisions between accelerated protons of energy $E_{p}$ and gas
protons, $J_{p}$ is the flux of accelerated protons, $E_{\pi}$ is
the energy of $\pi^{0}$ mesons; (ii) decay of $\pi^{0}$ resulting
in the gamma emission spectrum

\[
F\left(E_{\gamma}\right)=2\intop_{E_{\gamma}+m_{\pi}^{2}c^{4}/4E_{\gamma}}^{\infty}\frac{F_{pp}\left(E_{\pi}\right)}{\sqrt{E_{\pi}^{2}-m_{\pi}^{2}c^{4}}}dE_{\pi}
\]
where $m_{\pi}$ is the pion rest mass. 

The results are shown in Fig.\ref{fig:Gamma-emission-from}. The best
fit to the Fermi and Agile data is provided by a TP energy distribution
($\propto E^{-2}$) below $p_{br}\simeq7$GeV/c with the spectrum
steepening by exactly one power above it. The spectrum steepening
is perfectly consistent with the proton partial escape described above
(with no parameters involved) and shown in Fig.\ref{fig:Scattering-zone-of}.
For comparison, a weakly NL spectrum, shown in Fig.\ref{fig:Spectra-of-protons},
is also used for these calculations (dashed line in Fig.\ref{fig:Gamma-emission-from}),
but its fit would require a somewhat stronger break ($\Delta q\gtrsim1$)
or a low momentum cut-off, Fig.\ref{fig:Gamma-emission-from}, i.e.
at least one additional free parameter. We will discuss the options
in the next section.

\section{Summary and outlook\label{sec:Summary-and-outlook}}

To summarize the results, the mechanism for a break in the spectrum
of shock accelerated protons suggested in \cite{MDS05,MDS_11NatCo}
is in excellent agreement with the recent \cite{Abdo10W44full} Fermi
LAT and Agile \cite{AgileW44_11} observations of the SNR W44. The
observed gamma ray spectrum most likely results from the decay of
$\pi^{0}$-mesons which are born in $p-p$ collisions of shock accelerated
protons with an ambient dense gas. The parent proton spectrum is best
represented by a classical test particle power law $\propto E^{-2}$,
steepening to $E^{-3}$ at $E_{br}\approx7GeV$ due to deteriorated
particle confinement caused by the ion-neutral collisions and the
resultant Alfven wave evanescence. The position of the break momentum
in the particle spectrum may be estimated using eq.(\ref{eq:p1}),
or conversely, the combination of parameters involved in this estimate
can be inferred from the measured break momentum. The cut-off momentum
is not constrained in this scenario. 

An alternative explanation, based on a different mechanism of the
break, associated with the change of the particle transport in the
CR shock precursor \cite{MD06} is also possible but is less definitive
in the spectrum slope variation $\Delta q$ across the break (see
also \cite{Uchiyama10} for the most recent alternative suggestions).
In addition, the mechanism \cite{MD06} would imply a considerable
nonlinearity, i.e. a stronger CR shock precompression than that suggested
by the radio observation of accelerated electrons and the inferred
100 GeV proton upper cutoff (see below). Still alternatively, assuming
the ``environmental'' break mechanism is at work, i.e. $\Delta q=1$,
but the shock structure is somewhat modified, we arrive at the $E^{-1.75}$
spectrum below the break (as the radio observations may suggest for
the electrons), and $E^{-2.75}$ above the break. A fit to the data
is marginally possible, but it would require a relatively low cut-off
momentum at about $100$ GeV/c. This possibility may be supported
or ruled out once the data (upper limit) around this energy become
available. 

As we noted, particle escape from the MC can quench the acceleration
process \cite{DruryEscape11}. This would certainly be the case if
the MC were filling the entire shock precursor. However, MCs are known
to be clumpy \cite{CrutcherMC99,Chev03,Pariz04,Inoue12}, and fill
only a small fraction (< 1-2\%) of the precursor. In this case the
acceleration process continues largely unimpeded (apart from the spectrum
steepening) but the accelerated protons illuminate the 'cloudlets'
and make them visible in $\gamma$-rays due to the high density target
material. Another concern is a faint or even lacking x-ray emission
that seems to be inconsistent with shocks impacting dense surroundings.
This issue has been recently dealt with in, e.g., \cite{DruryHeating09,Inoue12}.
Large clumps survive the shock passage as it stalls inside them and
no strong heating occurs \cite{Inoue12}.

The most robust and attractive aspect of the suggested mechanism for
the spectral break is the exact $\Delta q=1$ variation of the spectral
index. Indeed, this change in the spectral slope is due to the reduction
of the number of degrees of freedom of particles caused by the resonant
wave evanescence and it does not depend on any parameters. In a combination
with the test particle regime operating below the break, which is
physically suggested by the low values of the break and upper cut-off
momenta, the mechanism provides a very good fit to the \emph{Fermi
}LAT and Agile data with no free parameters for the SNR W44 and probably
for W51C. From a number of physically different types of spectral
breaks suggested \cite{AharAt96,mdj02,MD06,OhiraMC_11}, namely the
current, ``environmental'' mechanism appears to be plausible where
a dense target gas is present which is also required for the efficient
$\pi^{0}$ production. However, observations of some other remnants
in the dense gas environments, such as W28 and IC443 \cite{AbdoW28_10,AbdoIC443_10}
indicate weaker breaks, $\Delta q=0.6-0.7$ which may either require
a different mechanism for the break or a narrower wave evanescence
gap $\Delta p=p_{2}-p_{1}$ (higher ionization rate). The predominance
of small clumps with $L_{{\rm c}}\ll L_{{\rm CR}}$ in a MC will also
reduce $\Delta q$.

Generally, spectral breaks offer a natural resolution to the well
known but puzzling trend of the \emph{nonlinear }(i.e. supposedly
improved) DSA theory to develop spectra which are considerably harder
than a simple test particle spectrum, thus becoming even less consistent
with the bulk of observations \cite{Gaisser98,Hillas05}. However,
the nonlinear spectrum -- i.e., diverging in energy-- exhausts the
shock energy available for the acceleration as the cut-off momentum
grows, so that a broken spectrum should form \cite{mdj02,MD06}. Broken
spectra are now commonly observed and the old paradigm of a single
power-law with an exponential upper cut-off is maladapted to the recent,
revolutionarily improved observations \cite{AbdoIC443_10,Abdo10W44full}.
Note, that the spectrum of the RX J1713.7-3946 \cite{Ahar06RXJ} is
also consistent with the environmental break mechanism presumably
operating in W44 surrounding but with a higher $p_{br}\sim10^{3}GeV/c$
and thus with stronger acceleration nonlinearity \cite{MDS05}. However,
it is difficult to make the case for hadronic origin of the gamma-ray
emission of the RX J1713.7-3946 \cite{AharNat04,Ahar06RXJ,WaxmanRXJ08}.
The fundamental role of the W44 remnant for the problem of CR origin
is that this particular remnant seems to rule out contaminating electron
emission due to Bremsstrahlung and inverse Compton scattering \cite{Abdo10W44full,Uchiyama10}
thus favoring the hadronic origin of the gamma emission and bolstering
the case for the SNR origin of galactic CRs
\begin{acknowledgments}
We would like to thank the anonymous referee for helpful suggestions.
Support by the Department of Energy, Grant No. DE-FG02-04ER54738 is gratefully
acknowledged. This work was also supported by the World Class Institute
(WCI) Program of the National Research Foundation of Korea (NRF) funded
by the Ministry of Education, Science and Technology of Korea (MEST)
(NRF Grant Number: WCI 2009-001)
\end{acknowledgments}
\clearpage

\bibliographystyle{apsrev}
\bibliography{C:/TeX/BIBS/dsa,C:/TeX/BIBS/DSAobs,C:/TeX/BIBS/MALKOV,C:/TeX/BIBS/PlasmaDSA,C:/TeX/BIBS/SNR,C:/TeX/BIBS/UHECR,C:/TeX/BIBS/textbooks,C:/TeX/BIBS/tokamak}

\end{document}